\documentclass [11pt]{article}
\usepackage{amsmath}
\usepackage{amsfonts}
\usepackage{epsfig}
\begin{document}
\def\b{\bar}
\def\d{\partial}
\def\D{\Delta}
\def\cD{{\cal D}}
\def\cK{{\cal K}}
\def\f{\varphi}
\def\g{\gamma}
\def\G{\Gamma}
\def\l{\lambda}
\def\L{\Lambda}
\def\M{{\Cal M}}
\def\m{\mu}
\def\n{\nu}
\def\p{\psi}
\def\q{\b q}
\def\r{\rho}
\def\t{\tau}
\def\x{\phi}
\def\X{\~\xi}
\def\~{\widetilde}
\def\h{\eta}
\def\bZ{\bar Z}
\def\cY{\bar Y}
\def\bY3{\bar Y_{,3}}
\def\Y3{Y_{,3}}
\def\z{\zeta}
\def\Z{{\b\zeta}}
\def\Y{{\bar Y}}
\def\cZ{{\bar Z}}
\def\`{\dot}
\def\be{\begin{equation}}
\def\ee{\end{equation}}
\def\bea{\begin{eqnarray}}
\def\eea{\end{eqnarray}}
\def\half{\frac{1}{2}}
\def\fn{\footnote}
\def\bh{black hole \ }
\def\cL{{\cal L}}
\def\cH{{\cal H}}
\def\cF{{\cal F}}
\def\cP{{\cal P}}
\def\cM{{\cal M}}
\def\ik{ik}
\def\mn{{\mu\nu}}
\def\a{\alpha}

\title{\textbf{Complex Structure of  the Four-Dimensional Kerr Geometry:
 Stringy System, Kerr Theorem, and Calabi-Yau Twofold}}
\author{Alexander Burinskii \\
\\
\emph{Theor.Phys. Lab., NSI, Russian Academy of Sciences,} \\
\emph{B. Tulskaya 52 Moscow 115191 Russia\footnote{E-mail address:
bur@ibrae.ac.ru} }}

\date{} \maketitle

\begin{abstract}
The 4d Kerr geometry displays many wonderful relations with
quantum world and, in particular, with  superstring theory.
The lightlike structure of fields near the Kerr singular ring
is similar to the structure of Sen solution for a closed
heterotic string.
Another string, open and complex, appears in the 
 complex representation of the Kerr geometry initiated by Lind and Newman.
Combination of these strings forms a membrane source of the Kerr
geometry which is parallel to the string/M-theory unification.
In this paper we give one more evidence of this relationship,
emergence of the Calabi-Yau twofold (K3 surface) in twistorial
structure of the Kerr geometry as a consequence of the Kerr theorem.
Finally, we indicate that the Kerr stringy system may correspond to
a complex embedding of the critical N=2 superstring.
\end{abstract}


\newpage

\section{Introduction}
The black hole solutions of diverse dimensions, represent now one
of the basic objects for study in superstring theory.  Recent
ideas and methods in the black hole physics are based on complex
analyticity and conformal field theory, which unifies the black
hole physics with superstring theory and physics of elementary
particles. The Kerr solution plays in this respect especial role.
Being  obtained as a metric of a "spinning mass" \cite{Kerr} with
angular momentum $J=m|a| ,$ the Kerr solution found basic
application as a metric of rotating black hole.
 In the four-dimensional Kerr solution, parameter $a= J/m$ is
 radius of the Kerr singular ring. For
$|a|<m ,$ the ring is covered by horizon, but for parameters of
the elementary particles $|a|>>m ,$ the black hole horizons
disappear, and the Kerr singular ring turns out to be naked.
Following the censorship principle, it should be covered by a
source. During four decades of investigations, structure of Kerr's
source was specified step by step. One of the earlier models was
the model of the Kerr ring as a closed string \cite{Bur0,IvBur}.
It has been obtained in \cite{BurSen} that structure of the fields
around the Kerr ring is similar to the structure of the heterotic
string in the solutions to low energy string
theory obtained by Sen \cite{KerSen}. However, the Kerr string is branch line of
the Kerr space-time into two sheets \cite{BurA}, and this bizarre
peculiarity created an alternative line of investigations of the
problem of Kerr's source,\cite{Keres,Isr,Lop,BurBag,BurSol}, which
led to conclusion that the source of the Kerr-Newman (KN) solution
should form a rigidly rotating membrane, or to be more precise, a
highly oblate ellipsoidal bubble with a flat vacuum interior
\cite{Lop,BurSol}.

 The charged KN solution \cite{KerNew} has found application as a
consistent with gravity classical model of spinning particle,
\cite{Car,DKS,Bur0,Isr,Lop,TN}, which has gyromagnetic ratio $g=2
,$ as that of the Dirac electron \cite{Car,DKS}, and also displays other
relationships  with the Dirac electron,\fn{In fact, the four observable
parameters of the electron: mass, spin, charge and magnetic moment indicate
unambiguously that the KN solution is to be the electron background geometry \cite{BurQ}.}
\cite{DirKer,BurAxi,BurQ,Beyond}, as well as the relationships
with twistor theory \cite{BurA,BurPreQ,BurTwi},   models of the soliton
\cite{BurBag,BurSol,Dym} and with basic structures of superstring
theory \cite{BurQ,BurTwi,IvBur,BurSen,BurCStr}.

In this note we consider complex structure of the Kerr geometry
\cite{BurCStr} and reveal one new evidence of its inherent
parallelism  with  superstring theory. Namely, we show the
presence of \emph{the Calabi-Yau twofold} (K3 surface) in complex
structure of the Kerr geometry, which appears as a consequence of
the Kerr theorem in the form of a quartic equation in the
projective twistor space $CP^3 .$ In section 2. we describe
briefly the real structure of the Kerr geometry and the Kerr
theorem, which determines Kerr's principal null congruence (PNC)
in twistor terms.

On the way to this principal result we meet a few important
intermediate complex and stringy structures. First of all it is
the complex Kerr geometry itself, which turns out to be  related
with the suggested by Appel in 1887 ``complex shift'', \cite{App},
and the reobtained and developed by Newman complex retarded-time
construction \cite{LinNew}. We describe them in section 3.

In section 4. we show that the source of the complex Kerr geometry
is to be an open complex string. It is closely linked with the old
remarks by Ooguri and Vafa, that the complex world lines (CWL)
parametrized by complex time parameter $\t=t+i\sigma ,$ are in
fact to be world-sheets of the complex string, \cite{OogVaf}. It
has been shown \cite{BurCStr} that the boundary conditions of the
complex Kerr string, generating the complex Kerr geometry, require
 orientifold structure of its world-sheet.

Finally, we observe that the structure of the membrane (vacuum
bubble \cite{BurSol}) source of the \emph{real} Kerr geometry
turns out to be parallel to formation of the membrane in the
superstring/M-theory unification. Namely, the  closed Kerr string
of the real Kerr geometry grows by extra world-sheet parameter of
the open complex Kerr string, \cite{BBS}, resulting in formation
of the Kerr bubble-membrane source, which is parallel to
construction of enhancon in string/M-theory \cite{LaJoh}.

This parallelism of the structure of Kerr geometry with basic
structures of the superstring theory, and in particular, the
inherent existence of the K3 surface, enforces us to draw a
parallel with the critical N=2 superstring theory \cite{GSW},
which describes a complex two-dimensional string (four real
dimensions) and is closely related with twistor theory too. This
allows us to suggest that the complex Kerr string represents an
embedding of the critical N=2 superstring theory into complex Kerr
geometry.

\section{Real structure of the KN geometry}

 KN metric is represented in the Kerr-Schild (KS) form \cite{DKS},
 \be g_\mn=\eta _\mn + 2h
e^3_\m e^3_\n \label{KSh} , \ee where $\eta_\mn$ is auxiliary
Minkowski background in Cartesian coordinates ${\rm x}= x^\m
=(t,x,y,z),$ \be h = P^2 \frac {mr-e^2/2} {r^2 + a^2 \cos^2
\theta}, \quad P=(1+Y\Y)/ \sqrt 2 ,  \label{h}\ee and $e^3 (\rm
x)$ is a tangent direction to a \emph{Principal Null Congruence
(PNC)}, which is determined by the form\fn{Here $ \z =
(x+iy)/\sqrt 2 ,\quad  \Z = (x-iy)/\sqrt 2 , u = (z + t)/\sqrt 2
,\quad v = (z - t)/\sqrt 2, $ are the null Cartesian coordinates,
$r, \theta, \phi $ are the Kerr oblate spheroidal coordinates, and
$Y (\rm x) =e^{i\phi} \tan \frac{\theta}{2} $ is a projective
angular coordinate.  The used signature is $(-+++) $.} \be e^3_\m
dx^\m =du + \bar Y d \zeta + Y d \bar\zeta - Y\bar Y dv ,
\label{e3} \ee via function $Y (\rm x),$ which is obtained from
\emph{the Kerr theorem},
\cite{DKS,KraSte,Pen,PenRin,BurKerr,Multiks}.

The PNC forms a
caustic at the Kerr singular ring, $r=\cos\theta=0 . $ As a
result, the KN metric (\ref{KSh}) and electromagnetic potential
 \be A_{\m} = -P^{-2} Re \frac {e} { (r+ia \cos \theta)}
e^3_\m \label{Amu} , \ee
are aligned with Kerr PNC and concentrate near the
 Kerr ring, forming a closed string -- waveguide for traveling electromagnetic waves
\cite{Bur0,IvBur,BurQ}.  Analysis of the Kerr-Sen solution to low
energy string theory \cite{KerSen} showed that similarity of the
Kerr ring with a closed strings is not only analogue, but it has
really the structure of a fundamental heterotic string
\cite{BurSen}. Along with this closed string, the KN geometry
contains also a \emph{complex open string}, \cite{BurCStr}, which
appears in the initiated by Newman complex representation of Kerr
geometry, \cite{LinNew}.  This string gives an extra dimension
$\theta$ to the stringy source ($\theta \in [0, \pi] $), resulting
in its extension to a membrane (bubble source \cite{Lop,BurSol}. A
superstring counterpart of this extension is a transfer from
superstring theory to $11$-dimensional $M$-theory and $M2$-brane,
\cite{BBS}.

\textbf{Kerr Theorem} determines the shear free null congruences
with tangent direction (\ref{e3}) by means of the solution $Y(\rm
x)$ of the equation \be F(T^A) =0 \label{F0KerrTeor}, \ee where
$F(T^A)$ is an arbitrary holomorphic function in the projective
twistor space with $CP^3$ coordinates  \be T^A= \{ Y, \quad \l ^1 = \z - Y v, \quad \l ^2
=u + Y \Z \} .\label{(TA)} \ee

Using the  Cartesian coordinates $x^\m ,$ one can rearrange
variables and reduce function $F(T^A)$ to the form $F(Y,x^\m),$
which allows one to get solution of the equation
(\ref{F0KerrTeor}) in the form $Y(x^\m).$

For the Kerr and KN solutions, the function $ F(Y,x^\m)$ turns out
to be quadratic in $Y,$ \be  F = A(x^\m)  Y^2 + B(x^\m) Y +
C(x^\m), \label{FKN} \ee and the equation (\ref{F0KerrTeor})
represents
 a \emph{quadric }in the projective twistor space $\bf CP^3 ,$
 with a non-degenerate determinant $\D = (B^2 - 4AC)^{1/2} $
which determines the complex radial distance \cite{BurKerr,BurNst}
\be \tilde r = - \D = -(B^2 - 4AC)^{1/2} . \label{trDet} \ee
 This case is explicitly resolved and yields two
   solutions \be Y^\pm (x^\m)= (- B \mp \tilde r )/2A, \label{Ypm}\ee
which allows one to restore two PNC by means of (\ref{e3}).

\noindent One can easily obtain from (\ref{FKN}) and (\ref{Ypm})
that the used in the metric (\ref{h}) and the em potential (\ref{Amu})
complex radial distance $\tilde r = r+ia\cos \theta$ may also be determined
from the Kerr generating function by the relation \be \tilde r = - dF/dY \label{tr} .\ee
Therefore, the Kerr singular ring, $\tilde r =0 ,$ is formed as a
caustic of the Kerr congruence, \be dF/dY=0 . \label{sing}\ee

\noindent As a consequence of  Vieta's formulas, the quadratic in $Y$
function (\ref{FKN}) may be expressed via the roots
$Y^\pm(x^\m)$ in the simple form \be
F(Y,x^\m)=A(Y-Y^+(x^\m))(Y-Y^-(x^\m)) \label{FYYpm} .\ee

\begin{figure}[ht]
\centerline{\epsfig{figure=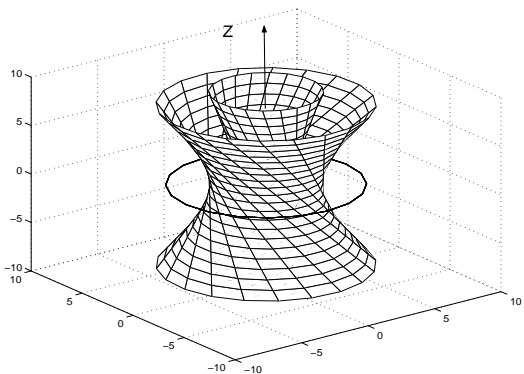,height=5cm,width=7cm}}
\caption{ Twistor null lines of the Kerr congruence are focused on
the Kerr singular ring, forming a twosheeted spacetime branched by
closed string.}
\end{figure}

\section{Complex Kerr geometry and the complex retarded-time construction}

KN solution was initially
obtained in \cite{KerNew} by a "complex trick" from the Kerr solution. One can see
 that the complex radial distance $\tilde r = r+ i a\cos\theta $ takes in
Cartesian coordinates the form
\be \tilde r =\sqrt {x^2+y^2
+(z+ia)^2}, \ee
and therefore, the scalar component of the vector potential (\ref{Amu}) may be
obtained from the Coulomb solution $\phi(\vec x) = e/r = e/\sqrt{x^2+y^2
+z^2} $ by a complex shift $z\to z+ia ,$ or by  the shift of its singular
point $\vec x_0 =(0,0,0)$ in complex region $\vec x_0 \to (0,0, -ia).$
This shift was first described in 1887 by Appel \cite{App}, who noticed
that the Coulomb solution, being invariant solution to the linear Laplace equation
by the real shifts $\vec x \to \vec x +\vec a ,$ should also be
invariant for the complex shift. In spite of triviality of this procedure from complex
point of view, it yields
very nontrivial consequences in the real section, in particular, the singular point
of the Coulomb solution $\vec x_0 =(0,0,0)$ turns into singular ring
$x^2+y^2+(z+ia)^2 =0 ,$ intersection of the sphere $x^2+y^2+z^2 =a^2 $ and plane $z=0,$
and the space turns out to be twosheeted, branching around this singular ring.

Lind and Newman showed, \cite{LinNew}, that the linearized KN solution corresponds to this
complex shift and may be generated by a complex source propagating along a
complex world line, and suggested a special complex retarded-time procedure
which generalizes the usual real retarded-time construction.
It has been shown later, \cite{BurKerr,BurNst}, that the complex retarded-time representation is exact, if the KN solution is presented in the Kerr-Schild form.
Therefore, the exact KN solution may be described as a field generated by a
\textbf{complex source propagating along complex world-line} \be
x_L^\m(\t_L) = x_0^\m (0) +  u^\m \t_L + \frac{ia}{2} \{ k^\m_L -
k^\m_R \} ,\label{cwL}\ee where $u^\m=(1,0,0,0), \quad
k_R=(1,0,0,-1), \quad k_L=(1,0,0,1).$ Index $L$ labels it as a
Left structure, and we should add a complex conjugate Right
structure \be x_R^\m(\t_R) = x_0^\m (0) +  u^\m \t_R -
\frac{ia}{2} \{ k^\m_L - k^\m_R \} .\label{cwR}\ee Therefore, from
complex point of view the Kerr and Schwarzschild geometries are
equivalent and differ only by their \emph{real slice}, which for
the Kerr solution goes aside of its center. Complex shift turns
the Schwarzschild radial directions $\vec n = \vec r /|r|$ into
twisted directions of the Kerr congruence, Fig.1.

\section{Complex Kerr's string } It was obtained
\cite{BurCStr,OogVaf} that the complex world line $x_0^\m (\t) ,$
parametrized by complex time $\t=t+i\sigma ,$ represents really a
two-dimensional surface which takes an intermediate position
between particle and string. The corresponding "hyperbolic string"
equation \cite{OogVaf},
 $\d_\t \d_{\bar\t}
x_0(t,\sigma) =0 ,$ yields the general solution \be x_0(t,\sigma)
= x_L(\t) + x_R(\bar\t) \ee as sum of the analytic and
anti-analytic modes $x_L(\t), \quad x_R(\bar\t),$ which are not
necessarily complex conjugate. For each real point $x^\m ,$ the
parameters $\t$ and $\bar\t$ should be determined by a complex
retarded-time construction. Complex source of the KN solution
corresponds to two \emph{straight} complex conjugate
world-lines,(\ref{cwL}),(\ref{cwR}). Contrary to the real case,
the complex retarded and advanced times $\t^\mp = t \mp \tilde r $ may
be determined by two different (Left or Right) complex null
planes, which are generators of the complex light cone. It yields
four different roots for the Left and Right complex structures
\cite{BurKerr,BurNst} \bea \t_L^\mp &=&
t \mp (r_L + ia\cos\theta_L) \label{Lretadv} \\
\t_R^\mp &=& t \mp (r_R + ia\cos\theta_R) \label{Rretadv}. \eea

The real slice condition determines relation $\sigma=a\cos \theta$
with  null directions  of the Kerr congruence $\theta \in [0, \pi]
,$ which puts restriction $\sigma \in [-a, a] $ indicating that
\emph{the complex string is open}, and  its endpoints $\sigma =
\pm a$ may be associated  with the Chan-Paton charges of a
quark-antiquark pair. In the real slice, the complex endpoints of
the string are mapped to the north and south twistor null lines,
$\theta =0,\pi ,$ see Fig.3.

\begin{figure}[ht]
\centerline{\epsfig{figure=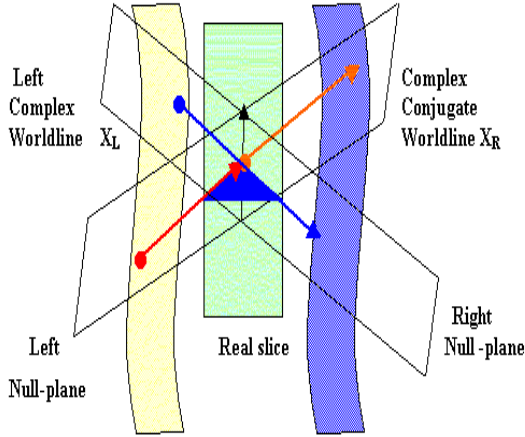,height=6.5cm,width=8.5cm}}
\caption{The complex conjugate Left and Right null planes generate
the Left and Right retarded and advanced roots.}
\end{figure}

\textbf{Orientifold.} The complex open string boundary
conditions \cite{BurCStr} require the \emph{worldsheet orientifold}
structure \cite{BBS,DHVW,HV,Hor,GSW} which
 turns the open string in a closed but
folded one. The world-sheet parity transformation $ \Omega: \sigma
\to - \sigma $ reverses orientation of the world sheet, and covers
it second time in mirror direction. Simultaneously, the Left and
Right modes are
 exchanged. \fn{Two oriented copies of the interval $\Sigma = [-a, a] ,$
$\Sigma^+ = [-a, a],$ and $ \Sigma^- = [-a, a]$ are joined,
forming a
 circle $ S^1 = \Sigma ^+\bigcup \Sigma ^-
,$ parametrized by $\theta ,$ and map $\theta \to \sigma=a\cos
\theta $ covers the world-sheet twice.} The projection $\Omega$ is
combined with space reflection $R: r\to - r ,$ resulting in
$R\Omega: \tilde r \to -\tilde r ,$ which relates the retarded and
advanced folds \be R\Omega:  \t^+ \to \t^- \label{POmt} ,\ee
preserving analyticity of the world-sheet.
\begin{figure}[ht]
\centerline{\epsfig{figure=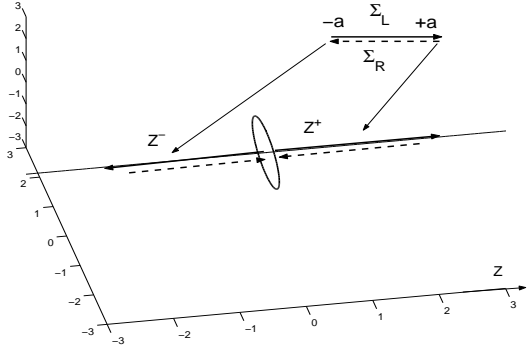,height=5cm,width=7cm}}
\caption{Ends of the open complex string, associated with quantum
numbers of quark-antiquark pair, are mapped onto the real
half-infinite $z^+, z^-$ axial strings. Dotted lines indicate
orientifold projection.}
\end{figure}
 The string modes $x_L(\t), \quad x_R(\bar\t),$ are extended on
the second half-cycle by the well known extrapolation,
\cite{BBS,GSW} \be x_L(\t^+) = x_R(\t^-); \quad x_R(\t^+) =
x_L(\t^-) , \label{orbi}\ee which forms the folded string, in which
the retarded and advanced modes are exchanged every half-cycle.

The real KN solution is generated by the straight complex world
line (CWL) (\ref{cwL}) and by its conjugate Right counterpart
(\ref{cwR}). By excitations of the complex string, the orientifold
condition (\ref{orbi}) becomes inconsistent with the complex
conjugation of the string ends, and \emph{ the world lines
$x_L(\t),$ and $x_R(\bar\t)$ should represent independent complex
sources}. The projection ${\cal T}=R\Omega $ sets parity between
the positive Kerr sheet determined
 by the Right retarded time and the negative sheet of the the Left
 advanced time. It allows one to escape the  anti-analytical Right
 complex structure,  replacing it by the Left advanced one, and
 the problem is reduced to self-interaction of the retarded and advanced
 sources determined by the time parameters $\t^\pm .$  For any
non-trivial (not straight) CWL, the Kerr theorem will generate
different congruences for $\t^+ ,$ and $\t^- .$  Each of these
sources produces a twosheeted Kerr-Schild geometry, and the formal
description of the resulting four-folded congruence should be
based on  the multi-particle Kerr-Schild solutions,
\cite{Multiks}.\fn{Physical motivation of such a splitting of the
sources is discussed in seminal paper by De Witt and Breme, \cite{DeWitBrem},
where authors introduce the similar `bi-tensor' fields, which are
predecessors  of the two-point Green's functions and Feynman propagator.
The problem of physical interpretation goes beyond frame of this paper,
and will be considered elsewhere.}
 The corresponding two-particle generating function
of the Kerr theorem will be \be F_2(T^A) = F_L(T^A) F_R(T^A)
,\label{F2} \ee where $F_L$ and $F_R$ are determined by
$x_L(\t^+),$ and $x_L(\t^-).$ The both factors are quadratic in
$T^A .$ The corresponding equation \be F_2(T^A)=0 \ee describes
\emph{a quartic in $CP^3 $} which is the well known Calabi-Yau
two-fold, \cite{BBS,GSW}. We arrive at the result that excitations
of the Kerr complex string generate a Calabi-Yau two-fold, or K3
surface, on the projective twistor space $CP^3 .$

\section{Outlook.} One sees that the Kerr-Schild geometry displays
striking parallelism with basic structures of  superstring theory.
However, our principal result in this paper is the presence of
inherent Calabi-Yau twofold in the complex twistorial structure of
the Kerr geometry. In the recent paper \cite{BurQ} we argued that
it is not accidental, because gravity is a fundamental part of the
superstring theory. However the  Kerr-Schild gravity,
being based on twistor theory,  displays also some inherent relationships with
superstring theory.

In many respects the Kerr-Schild gravity resembles the
twistor-string theory, \cite{Nair,Wit,BurTwi}, which is also
four-dimensional, based on twistors and related with experimental
particle physics. On the other hand, the complex Kerr string has mach in
common  with the N=2 superstring \cite{OogVaf,GSW,DAddaLiz}. It is
also related with twistors and has the complex critical dimension
two which corresponds to four real dimensions and indicated that
N=2 superstring may lead to four-dimensions. However, signature of
the N=2 string may only be (2,2) or (4,0), which caused the obstacles
for embedding of this string in the space-times with minkowskian
signature. Up to our knowledge, this trouble was not resolved so far, and
the initially enormous interest to N=2 string seems to be dampened. Meanwhile,
embedding of the N=2 string in the complexified Kerr geometry is almost
trivial task. It hints that stringlike structures
of the real and complex Kerr geometry are not simply analogues, but
 reflect the underlying dynamics of the N=2 superstring theory,

In the same time, along with wonderful
parallelism, the stringy system of the four-dimensional KN geometry
displays  very essential peculiarities, which make it closer to particle physics.

\begin{itemize}
\item The spin/mass ratio of the spinning particles is extremely high which leads to the over-rotating BH geometry without horizons and offers a new application, next to
    traditional attention of superstring theory to quantum black holes.

\item  The supplementary Kaluza-Klein space is absent, and the role of
compactification circle is played by the naked Kerr singular ring with traveling waves,
which realizes a "compactification without compactification", \cite{BurQ}.

\item The lightlike twistorial rays are tangent to the  Kerr singular ring, indicating that the Kerr ring is the lightlike string, and it may play the role of  DLCQ circle
of M-theory, \cite{Suss}.

\item Consistency of the KN solution with gravitational background  of the electron,
\cite{Isr,Lop,Car,BurQ}, shows that the 4d Kerr characteristic length of
the Kerr ring,  $a=J/m ,$ corresponds to the Compton scale of spinning particles.

\end{itemize}

The considered stringy structures of the real and complex Kerr geometry set a parallelism between the
4d Kerr geometry and superstring theory, indicating that complexification of
the Kerr geometry may serve an alternative to traditional compactification of higher dimensions.

{\bf Acknowledgements.} Author thanks T. Nieuwenhuizen for permanent
interest to this work and useful conversations, and also very
thankful to Dirk Bouwmeester for  invitation to Leiden University,
where this work was finally crystalized in the process of the numerous
discussions with him and members of his group: Jan W. Dalhuisen,
V.A.L. Thompson and J.M.S. Swearngin. Author is also very
thankful to Laur Jh\"arv for very useful discussion at
the Tallinn conference "3Quatum" and for the given reference to the
related paper \cite{LaJoh}.

\end{document}